\documentclass[journal,twocolumn]{IEEEtran}
\usepackage{epsfig,makeidx,color}
\usepackage{amsmath,amssymb,bbm}
\usepackage{cite,graphicx,lipsum}
\usepackage{enumerate}
\newcommand{\indicator}{\mathbbm{1}}

\def\defhR{\mbox{\footnotesize 
$\begin{array}{c} \cH_1 \cr > \cr < \cr \cH_0 \end{array}$}}

\def\cH{{\cal H}}
\def\cK{{\cal K}}

\def\cQ{{\cal Q}}

\def\rH{{\rm H}}
\def\rT{{\rm T}}

\def\uC{{\mathbb C}}
\def\uE{{\mathbb E}}
\def\uP{{\mathbb P}}

\newtheorem{mylemma}{\bf Lemma} 

\def\be{ \begin{equation} }
\def\ee{ \end{equation} }
\def\bea{ \begin{eqnarray} }
\def\eea{ \end{eqnarray} }

\def\bx{{\bf x}}
\def\by{{\bf y}}
\def\bc{{\bf c}}

\def\ba{{\bf a}}
\def\br{{\bf r}}

\def\bn{{\bf n}}

\def\bp{{\bf p}}

\def\bC{{\bf C}}

\def\bI{{\bf I}}

\def\bR{{\bf R}}

\def\b0{{\bf 0}}

\def\cA{{\cal A}}
\def\cC{{\cal C}}

\def\cQ{{\cal Q}}

\def\cN{{\cal N}}

\ifCLASSOPTIONonecolumn
  \newcommand{\figwidth}{0.50\columnwidth}
  
\else
  \newcommand{\figwidth}{0.90\columnwidth}
  
\fi

\begin{document}

\title{Performance Analysis of 2-Step Random Access with CDMA
in Machine-Type Communication}

\author{Jinho Choi\\
\thanks{The author is with
the School of Information Technology,
Deakin University, Geelong, VIC 3220, Australia
(e-mail: jinho.choi@deakin.edu.au).
This research was supported
by the Australian Government through the Australian Research
Council's Discovery Projects funding scheme (DP200100391).}}

\maketitle
\begin{abstract}
There is a growing interest in
the transition from 4-step random access to 2-step random access 
in machine-type communication (MTC),
since 2-step random access is well-suited to
short message delivery
in various Internet of Things (IoT) applications.
In this paper, we study a 2-step random access
approach that uses
code division multiple access (CDMA) to form
multiple channels for data packet transmissions
with a spreading factor less than the number of channels.
As a result, the length of data transmission phase
in 2-step random access
can be shorter at the cost of multiuser interference.
To see how the decrease of the length of data transmission phase
can improve the spectral efficiency, we derive
the throughput as well as the spectral efficiency. From the 
results of analysis, we can show that the 2-step CDMA-based
random access approach can have a higher spectral efficiency
than conventional 2-step approach with orthogonal
channel allocations, which 
means that the performance of MTC 
can be improved by successfully transmitting more 
data packets per unit time using CDMA.
This is also confirmed by simulation results.
\end{abstract}


\ifCLASSOPTIONonecolumn
\baselineskip 26pt
\fi

\section{Introduction}

The Internet of Things (IoT) has been
extensively studied 
to connect a large number of devices 
for various applications including
smart home, smart cities, smart manufacturing, and so on
\cite{Gubbi13} \cite{Kim16}.
To support the connectivity for various IoT applications
over a wide area,
a number of different options can be considered
\cite{Ding_20Access}. Among those, 
in cellular IoT, machine-type communication (MTC)
\cite{3GPP_MTC} \cite{3GPP_NBIoT} 
plays a crucial role in providing the connectivity for IoT
devices and sensors through cellular systems
\cite{Shar15} \cite{Bockelmann16}.
Due to sparse activity of those devices and sensors,
MTC is usually based on random access 
to keep signaling overhead low 
\cite{3GPP_MTC} \cite{3GPP_NBIoT} 
\cite{Galinina13} \cite{Choi16}.

In MTC, a fraction of devices are active
at a time and use random access 
to establish connections to transmit their data 
(e.g., random
access channel (RACH) procedure 
in the long-term evolution advanced (LTE-A) systems \cite{3GPP_MTC}).
For random access, a common pool of preambles
is used \cite{3GPP_MTC} \cite{3GPP_NBIoT}.
A device that has data packets to transmit, which is called
an active device, randomly chooses 
a preamble from the pool and transmits it to a
base station (BS)
(through physical random access channel (PRACH) in RACH procedure)
which is the first step of a handshaking process
to establish connection in most MTC schemes (e.g.,
\cite{3GPP_MTC}).
Due to multiple active devices that choose
the same preamble, there exist preamble collisions and
this step can be seen as contention-based access.
There are 3 more steps 
in the handshaking process
to allocate dedicated uplink (data) channels
(which are physical uplink shared channel (PUSCH)
in RACH procedure) 
to active devices so that they can transmit their data packets
to the BS, which is the third step. 
Note that the 2nd and 4th steps are feedback
transmissions from the BS to devices.
The resulting random access approaches are called
4-step random access as there are 4 main steps.

Provided that messages transmitted from 
devices are short, 
to improve efficiency, 4-step random access
can be replaced with 2-step random access
\cite{Bockelmann18},
which is also referred to as grant-free random access
or compressive random access (CRA)
\cite{Schepker13} \cite{Wunder14}
\cite{ChoiYu17} \cite{Choi17IoT} \cite{Abebe17}.
The first step consists of two stages, 
where an active device transmits a preamble and then
a data packet. In other words,
the first and third steps of 4-step random access
are combined into the first step of 2-step random access,
while the second step is to send feedback
signals from the BS to devices.
In general, 2-step random access can be seen as a
multichannel random access scheme,
where each (multiple access) channel can be characterized by
a preamble in the first stage of the first step.
In the second stage of the first step, the preamble can be used
as a spreading sequence to transmit
data packet, which results in
code division multiple access (CDMA) based random
access \cite{Zhu11} \cite{Applebaum12}.
As shown in \cite{Abebe17}, 
it is also possible to use different spreading code (but associated
with the preamble in the first stage) for spreading
in the second stage.

If a BS is equipped with multiple antennas,
based on the notion of massive multiple input multiple output (MIMO)
\cite{Marzetta10}, without spreading,
the BS is able to decode data packets transmitted by
multiple active devices in a shared channel simultaneously
\cite{deC17}  \cite{Senel17} \cite{Liu18} \cite{Liu18a}.
Due to unbounded capacity of massive MIMO
\cite{Bjornson18}, 
it seems that massive MIMO is a solution to massive MTC. However, 
the cost of BSs with a large number of antenna elements
may be high. Thus, for small
cells, a BS equipped with single or a few
antennas could be practical for MTC. As a result,
in 2-step random access,
for data transmissions in the second stage,
CDMA or any orthogonal multiple access scheme
such as time division multiple access (TDMA) has to be employed.
When TDMA is used to support multiple data packet transmissions,
the data transmission phase is to be divided
into the same number of blocks as the number of preambles
so that an active device that chooses the $l$th
preamble is to send its data packet through
the $l$th block in the 2nd stage.
However, this may result in waste of resources
if the number of active devices
is less than that of blocks.
To avoid it, in \cite{Choi20b}, 
the number of blocks is dynamically decided
by the number of active devices.
However, since this approach requires feedback after
the preamble transmission,
it is not qualified as a 2-step random access approach.

In this paper, we consider a 2-step random access scheme
where CDMA is used for transmitting
data packets from multiple active devices.
In addition, a set of orthogonal preambles is assumed
so that the 
channel state information (CSI) of each active device
can be estimated without interference if there is no preamble collision.
As in \cite{Abebe17}, it is
assumed that each preamble is associated
with a different spreading sequence\footnote{It is noteworthy 
that the system model 
of this paper is similar to that in \cite{Abebe17}, while 
we present a detailed throughput analysis that was not addressed
in \cite{Abebe17}.}.
In particular, the case that the length of
spreading sequence is shorter than that of preambles
is considered. As a result, 
the length of the second stage can be shorter,
which can result in an improved spectral efficiency.
However, unlike preambles, spreading sequences are not orthogonal
(as their length is shorter than the number of them)
and there exists multiuser interference.
To mitigate  multiuser interference,
multiuser detection \cite{VerduBook} \cite{ChoiJBook2}
can be used under certain conditions.
Since the performance depends on key parameters
(e.g., the number of active
devices, the number of preambles, and the length
of spreading sequences),
the throughput and spectral efficiency are 
derived in terms of those key parameters,
which are the main contributions of the paper.
From the analysis,
we can show that 2-step random
access with CDMA can outperform 2-step random access with 
orthogonal channel allocations, e.g., TDMA.

Note that preambles are used not only for the channel estimation,
but also for user activity detection. Thus, preamble
design is important, although we do
not consider it (we only use a set of orthogonal
preambles). Thus, there are various approaches for 
preamble design. For example,
in \cite{Cheng20}, a preamble design
is considered to allow efficient user activity detection.
In \cite{Choi20a}, an approach to effectively
increase the number of preambles is studied.

The rest of the paper is organized as follows.
In Section~\ref{S:SM}, we present the system model for
2-step random access in MTC.
CDMA is considered in 
Section~\ref{S:CDMA} to form multiple channels for 
data packet transmissions by multiple devices simultaneously.  
We briefly discuss preamble detection with power control
in Section~\ref{S:PD} and multiuser detection
in Section~\ref{S:MUD}.
The main contribution of the paper is presented in
Section~\ref{S:Thp}, 
where the throughput and spectral efficiency are derived 
for 2-step random access with CDMA.
Simulation
results are presented in Section~\ref{S:Sim},
and the paper is concluded with some remarks in Section~\ref{S:Conc}.

{\it Notation}:
Matrices and vectors are denoted by upper- and lower-case
boldface letters, respectively.
The superscripts $\rT$ and $\rH$
denote the transpose and complex conjugate, respectively.
The support of a vector is denoted by ${\rm supp} (\bx)$
(which is the number of the non-zero elements of $\bx$).
$\uE[\cdot]$
and ${\rm Var}(\cdot)$
denote the statistical expectation and variance, respectively.
$\cC \cN(\ba, \bR)$
represents the distribution of
circularly symmetric complex Gaussian (CSCG)
random vectors with mean vector $\ba$ and
covariance matrix $\bR$.

\section{System Model}	\label{S:SM}

In this section, we consider a two-step random access system
that consists of one BS and a number
of devices for MTC.

\begin{figure}[thb]
\begin{center}
\includegraphics[width=\figwidth]{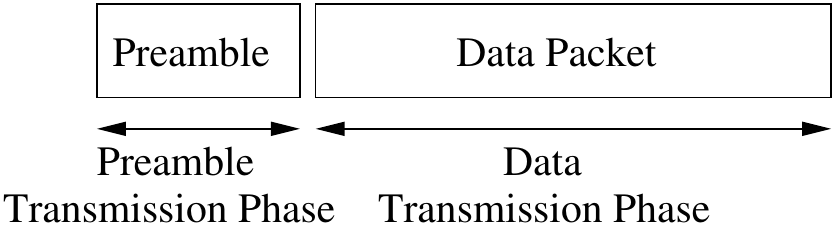}
\end{center}
\caption{Two phases
(i.e., preamble transmission and data transmission
phases) for a 2-step random access scheme.}
        \label{Fig:two_phase}
\end{figure}

Compared with 4-step random access approaches
\cite{3GPP_MTC} \cite{3GPP_NBIoT},
2-step random access, which is also referred
to as grant-free random access, can be more efficient
due to low signaling overhead when devices' have short messages.
In 2-step random access, the first step consists of
preamble and data transmission phases
as illustrated in Fig~\ref{Fig:two_phase}.
An active device is to choose
a preamble from a pool of $L$ pre-determined preambles,
denoted by $\{\bp_1, \ldots, \bp_L\}$,
and transmit it in the preamble transmission phase.
Throughout the paper, we assume that preambles
are orthonormal, i.e.,
$$
\bp_l^\rH \bp_{l^\prime} = \delta_{l, l^\prime},
$$
and the length of $\bp_l$ is $L$, i.e., $\bp_l \in \uC^L$.
Here, $\uC^n$ represents the 
$n$-dimensional complex coordinate space.

A data packet is then transmitted in the data transmission phase.
It is assumed that a time slot is used to transmit a preamble
and a data packet. Since all the devices are synchronized in MTC,
it is expected that the length of data packet is the same
for all devices (or the length of data transmission phase
is decided by the maximum length of data packet among
all the devices).

The BS receives the signals from multiple active devices and performs
preamble detection. That is, the BS is to detect transmitted preambles
among the $L$ preambles. For each detected preamble,
the BS can estimate the CSI of the associated active device
and use it to decode the following data packet (via coherent detection).

In the second step, the BS broadcasts its decoding results 
to the devices with the feedback signal of
acknowledgment (ACK) or negative acknowledgment (NACK).

Since the number of preambles, $L$, is limited,
multiple active devices can choose the same preamble,
which results in preamble collision.
For a transmitted preamble  by multiple active devices,
the estimate of CSI becomes a noisy superposition of
multiple channel coefficients. Thus,
decoding associated with the transmitted preamble
will fail. In other words, preamble collision leads to
NACK to all the devices associated preamble collision.

\section{Data Packet Transmissions in Two-Step Random Access}	\label{S:CDMA}

If there are multiple active devices,
they send data packets simultaneously
during the data transmission phase in 2-step random access.
Thus, there should be multiple channels.
In this section, we discuss two different approaches
to form multiple channels for data transmission
when the BS does not have a large number of antennas
(we also briefly explain the case when the BS has a large
number of antennas later in this section).

\subsection{Time Division Multiple Access}


In 2-step random access, the maximum
number of channels for data packet transmissions is
$L$, in which each channel is associated with one preamble.
That is, an active device that chooses and transmits preamble $l$
is to send its data packet in data channel $l$.
TDMA can be used for orthogonal channel allocations for data transmissions.
In this case, the sub-slot for the data transmission phase
is divided into $L$ time blocks 
and the resulting approach is referred to as
the TDMA-based approach for convenience.
For convenience, denote by $D$ the length of data packet
of each active device for short message delivery.
Then, the length of the data transmission phase becomes $D L$.

\subsection{Code Division Multiple Access}

There is a drawback of 
the TDMA-based approach. Suppose that the $l$th preamble
is not chosen by any active devices.
Then, clearly, the $l$th block becomes empty, which can result
in a low spectral efficiency.
Thus, in this paper, we study an approach
based on CDMA for data packet transmissions,
which is similar to the approach in \cite{Abebe17}.

Let $\bc_l$ denote the spreading sequence of length $N$
associated with the $l$th preamble. 
Here, $N$ is also called the spreading factor \cite{ViterbiBook}.
Unlike preambles, it is assumed that the $\bc_l$'s are not orthogonal
as $N < L$ (i.e., the length of spreading
sequences is shorter than that of preambles).
The resulting approach is referred to as the CDMA-based approach.
Since $N < L$, the length of the data transmission
phase of the CDMA-based approach
is shorter than that of the TDMA-based approach,
which can improve the spectral efficiency
at the cost of multiuser interference.

\begin{figure}[thb]
\begin{center}
\includegraphics[width=\figwidth]{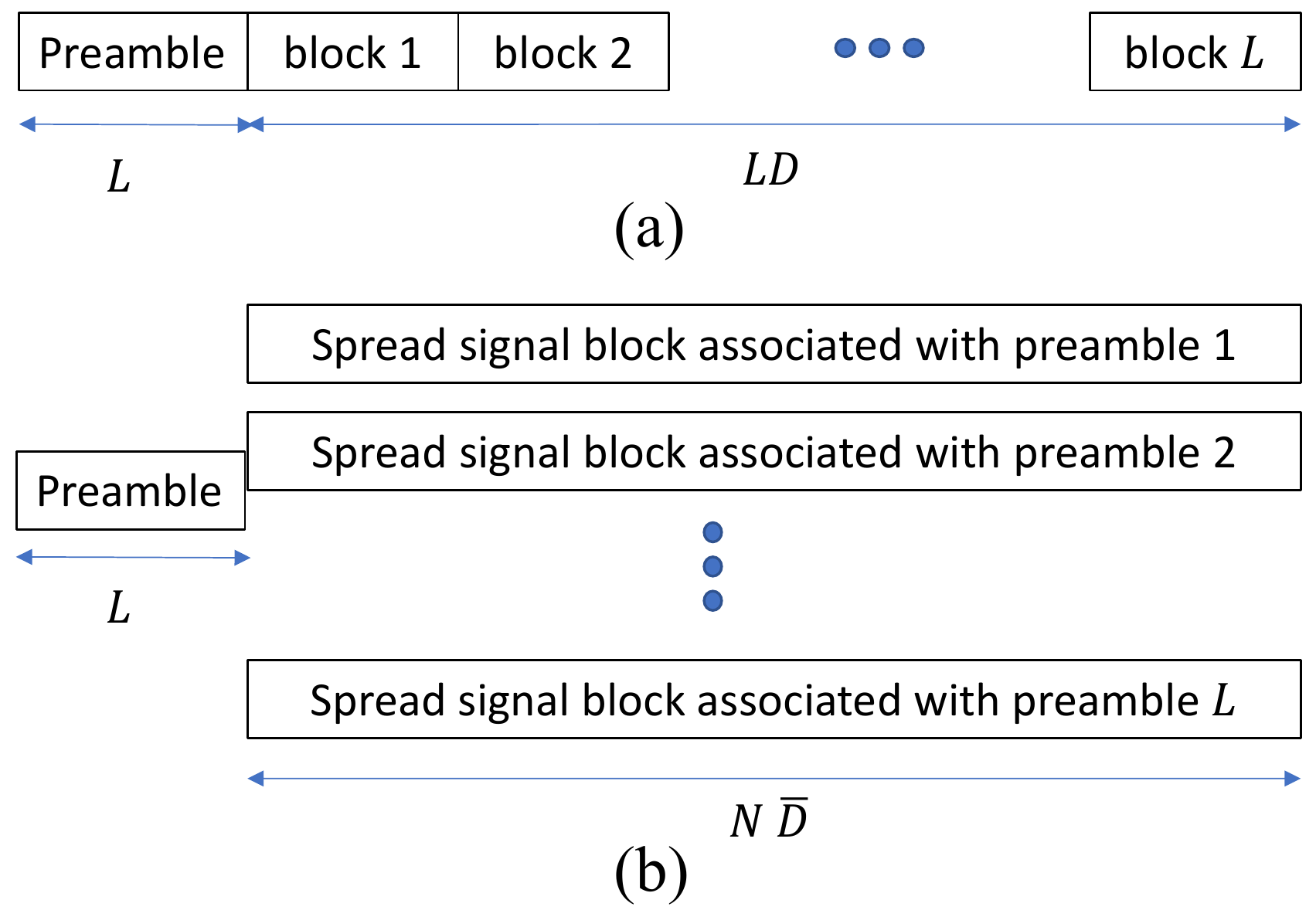}
\end{center}
\caption{Two different structures for the data
transmission phase: (a) TDMA-based approach 
with the data transmission phase of length $DL$;
(b) the CDMA-based approach
with the data transmission phase of length $\bar DN$.}
        \label{Fig:Fig2}
\end{figure}

For convenience, however, we assume that the length of slot 
is the same for both TDMA and CDMA, 
which is denoted by $T_{\rm slot}$.
In this case,
since $N < L$, the number of data symbols in data
transmissions becomes $\bar D = \frac{LD}{N}$
as illustrated in Fig.~\ref{Fig:Fig2}.
Here, we have $T_{\rm slot} = LD +L = L + N \bar D$.

There are a few remarks as follows.
\begin{itemize}
\item If $N = L$, we can use orthogonal
spreading sequences (i.e., $\bc_l = \bp_l$ for all $l$). However,
the resulting approach is equivalent to
the TDMA-based approach (as each spreading
sequence can be seen as an orthogonal block
in the time domain). Thus, for the CDMA-based approach,
we only need to consider $N < L$.

\item With the same length of slots, the number of data symbols
per slot in the CDMA-based approach, $\bar D$,
is larger than that the TDMA-based approach, $D$, as long as $N < L$.
In addition, as $\frac{L}{N}$ decreases, 
the CDMA-based approach becomes more efficient
than the TDMA-based approach in terms of spectral efficiency.
However, 
since decoding fails if $N$ is smaller than the number of active devices
in the CDMA-based approach (as will be shown in 
Section~\ref{S:MUD}),
$N$ cannot be arbitrarily small. Thus,
throughput will be analyzed in Section~\ref{S:Thp},
to see the impact of $N$ on the performance.

\item The length of slot
cannot be long for coherent detection. 
As a result, 2-step random access would be preferable
only when the length of message (i.e., $D$
or $\bar D$) is sufficiently short
and the number of devices is not too large (to limit $L$).
Thus, to support a large number of devices,
multiple 2-step random access schemes 
should run in parallel over multiple resource blocks.
\end{itemize}

\subsection{With Massive MIMO}

If the BS is equipped with a large antenna array,
the notion of MIMO can be exploited \cite{Marzetta10}.
In particular, the BS can estimate the channel vectors
using the bank of the correlators with $L$ preambles.
Provided that there is one active device for a given 
preamble, the output of the correlator can
provide an estimate of the channel vector of the active device.
As the number of antennas increases,
it is known that the channel vectors of different users or devices
become asymptotically orthogonal. Thus,
although all the active devices transmit their data
packets through \emph{one} data channel,
the BS is able to decode them by exploiting a high 
spatial selectivity
\cite{deC17}  \cite{Senel17} \cite{Liu18} \cite{Liu18a}.
In this case, the length of data transmission phase becomes $D$. 
Certainly, two-step random access with massive MIMO
has a high spectral efficiency
at the cost of a large antenna array, which is often infeasible
for small BSs that support relatively small areas (i.e., small
cells).

%
%

\section{Preamble Detection and Performance Analysis}	\label{S:PD}

In this section, we discuss preamble detection
and show that reliable preamble detection
can be achieved under power control.
In the rest of the paper, we assume that
the BS is equipped with a single antenna. Thus, no spatial 
diversity is to be exploited at the BS.

\subsection{Preamble Detection with Power Control}	\label{SS:PD}

Suppose that there are $K$ active devices.
Let $\cK_l$ denote the index set of the
active devices that choose preamble $l$.
In addition, denote by $h_k \in \uC$ the channel coefficient\footnote{Throughout
the paper, we consider a narrowband system. As a result,
the channel from each device to the BS
can be seen as a flat fading channel.}
from the $k$th active device to the BS.
Then, the received signal at the BS
during the preamble transmission phase is given by
\begin{align}
\by 
& = \sum_{1=1}^L \bp_l \sum_{k \in \cK_l} 
h_k \sqrt{P_k} + \bn_{\rm 1} \cr
& = \sum_{k=1}^K \bp_{l(k)} 
h_k \sqrt{P_k} + \bn_{\rm 1} \in \uC^L ,
\end{align}
where $P_k$ stands for the transmit power of 
active device $k$, $l(k)$
is the index of the preamble that is
chosen by active device $k$,
and $\bn_1 \sim \cC \cN(0, N_0 \bI)$
is the background noise vector.

Throughout the paper, it is assumed that
the transmit power is decided according to the following power control
policy:
\be
|h_k \sqrt{P_k}|^2 =
|h_k|^2 P_k = P_{\rm rx},
	\label{EQ:hPP}
\ee
where $P_{\rm rx}$ represents the target receive power at the BS.
In order to allow the power control in 
\eqref{EQ:hPP},
it has to be assumed that any active device knows
its channel coefficient $h_k$ prior to preamble transmission
so that its power can be decided according to \eqref{EQ:hPP}.
To this end, we assume that the system is based
on time division duplexing (TDD) so that
active devices can estimate their channel coefficients
using the channel reciprocity \cite{Smith04}
when the BS sends a beacon or pilot signal that
is to be transmitted periodically.

Furthermore, it is assumed that the coherence
time is sufficiently long so that $h_k$ remains
unchanged over the duration of slot, which might be a reasonably
assumption when devices are not mobile.

Note that if the magnitude of the channel coefficient is not
sufficiently large and the transmit power can exceed
the maximum transmit power\footnote{Note
that the norm of preambles is normalized,
i.e., $||\bp_l|| = 1$, which means that
$P_k = P_k ||\bp_{l(k)}||^2$ is the total energy to 
transmit a preamble by an active device.
Thus, although actual transmit power is limited,
$P_k$ can be high if a longer preamble is used.},
denoted by $P_{\rm max}$, i.e.,
$$
P_k = \frac{P_{\rm rx}}{|h_k|^2} > P_{\rm max},
$$
we assume that the device cannot transmit signals.
If such a device transmits a preamble,
the received signal does not have a sufficiently high SINR
to allow reliable preamble detection, while it becomes an interfering
signal to another active device that happens to choose
the same preamble. Thus, the device that cannot meet
\eqref{EQ:hPP} should not be active.

For convenience, let $h_k = |h_k| e^{j \theta_k}$,
where $\theta_k$ is the phase of $h_k$.
In this case, from \eqref{EQ:hPP},
we have $h_k \sqrt{P_k} = \sqrt{P_{\rm rx}} e^{j \theta_k}$.
Since the preambles are orthonormal,
the bank of $L$ correlators can be used to detect
transmitted preambles. The output of the $l$th correlator is
given by
\begin{align}
z_l 
& = \bp_l^\rH \by 
=  \sum_{k \in \cK_l} h_k \sqrt{P_k} + \bp_l^\rH \bn_1 \cr
& = \sqrt{P_{\rm rx}} 
a_l +  n_l, 
	\label{EQ:un}
\end{align}
where $a_l = \sum_{k \in \cK_l} e^{j \theta_k}$
and $n_l = \bp_l^\rH \bn_1$.
Thanks to orthonormal preambles,
we have $n_l \sim \cC \cN(0, N_0)$.

Provided that devices can estimate
$h_k$ precisely based on the channel reciprocity,
it is possible that 
each active device can compensate the phase of
the signal to be transmitted. For example,
the signal that active device $k$ transmits
is $\sqrt{P_k} e^{-j \theta} \bp_{l(k)}$.
Then,  $a_l$ in \eqref{EQ:un} with phase compensation becomes
\be
a_l =  \sum_{k \in \cK_l} e^{j \theta_k}e^{-j \theta_k}
 = |\cK_l|.
	\label{EQ:al_with}
\ee

For convenience, define the index set of 
transmitted preambles as
$$
\cA = \{l\,\bigl|\, |\cK_l| \ge 1\},
$$
Due to the possibility that the same preamble can be chosen
by multiple active devices, we have
\be
Q = |\cA| \le \min\{L, K\}.
\ee
For convenience,
$\bar l (q)$
denotes the $q$th element of $\cA$
or the index of the $q$th detected preamble in ascending order,
i.e., $\cA = \{\bar l(1), \ldots, \bar l(Q)\}$,
where $\bar l (1) < \ldots < \bar l (Q)$.
For example, suppose that $L = 4$ and $K = 3$.
The first and second active devices choose preamble 4,
and the third active device chooses preamble 1.
Then, at the BS, if all the transmitted preambles are correctly
detected, we have
$Q = 2$ and $\cA = \{1, 4\}$. That is, $\bar l (1) = 1$
and $\bar l (2) = 4$,
while $l(1) = l(2) = 4$ and $l(3) = 1$.
Note that if there is no preamble collision (which requires
that $K \le L$), we have $Q = K$ and 
there is a one-to-one relationship
between $\{\bar l(q) \}$ and $\{l(k)\}$.

The ideal outcome of the preamble detection
is to obtain $\cA$.
From $\{z_l\}$,
hypothesis testing can be carried out
to estimate $\cA$ 
with and without phase compensation at devices.

\subsection{Hypothesis Testing and Performance Analysis}

\subsubsection{With Phase Compensation}

With phase compensation at active devices,
$a_l$ is given in \eqref{EQ:al_with}.
Thus, the test statistic to detect the presence of 
preamble $l$ in $z_l$ becomes $\Re(z_l)$,
which can be written as
\be
\Re(z_l) = \sqrt{P_{\rm rx}} |\cK_l| + \Re(n_l),
\ee
where $\Re(n_l) \sim \cN \left(0, \frac{N_0}{2} \right)$.

Letting $\cH_0$ and $\cH_1$ represent the hypotheses of
the absence and presence of preamble $l$, respectively,
the following hypothesis testing can be 
considered:
\be
\Re(z_l) \defhR \tau,
	\label{EQ:HP}
\ee
where $\tau > 0$ is a decision threshold.
Under $\cH_1$, $\min |\cK_l|  = 1$. Thus,
we only consider the case that $|\cK_l| = 1$ for $\cH_1$.
Then, from \eqref{EQ:HP},
the probability of missed detection (MD) is given by
\begin{align}
\uP_{\rm MD} 
& = \Pr\left(\Re(z_l) \le \tau \,\bigl|\, |\cK_l| = 1 \right) \cr
& = \cQ
\left( \sqrt{\frac{2}{N_0}}
\left( \sqrt{P_{\rm rx}} - \tau \right) \right).
\end{align}
On the other hand, the probability of false alarm (FA)
is 
\begin{align}
\uP_{\rm FA} 
= \Pr\left(\Re(z_l) \ge \tau \,\bigl|\, |\cK_l| = 0 \right) 
= \cQ \left( \sqrt{\frac{2}{N_0} } \tau \right).
\end{align}
If $\tau = \frac{\sqrt{P_{\rm rx}}}{2}$,
we have
$\uP_{\rm MD} 
=\uP_{\rm FA}  = 
\cQ \left( \sqrt{\frac{P_{\rm rx}}{2 N_0} } \right)$.
Clearly, with a sufficiently high
signal to noise ratio (SNR), $\frac{P_{\rm rx}}{N_0}$, 
the probabilities of MD
and FA become low,
which implies that the 
preamble detection can be carried out at the BS reliably
or $\cA$ can be obtained with a high probability.
For example, if the SNR is 10 dB,
$\cQ\left(\sqrt{\frac{\rm SNR}{2}} \right) = 0.0127$.

\subsubsection{Without Phase Compensation}

In TDD mode, due to the channel reciprocity,
it is assumed that an active device can estimate the 
channel coefficient, $h_k$, for phase compensation. 
However, the radio frequency (RF) circuits on the BS 
differ from those on devices. Thus,
the phase of $h_k$ may not be precisely estimated
without calibration \cite{Guillaud13},
while the power control in \eqref{EQ:hPP} can be carried out
reliably as it is based on the magnitude, i.e., $|h_k|$.
Consequently, the phase compensation
may not be affordable at low-cost devices,
and we now consider the hypothesis testing
without phase compensation.

\begin{figure}[thb]
\begin{center}
\includegraphics[width=\figwidth]{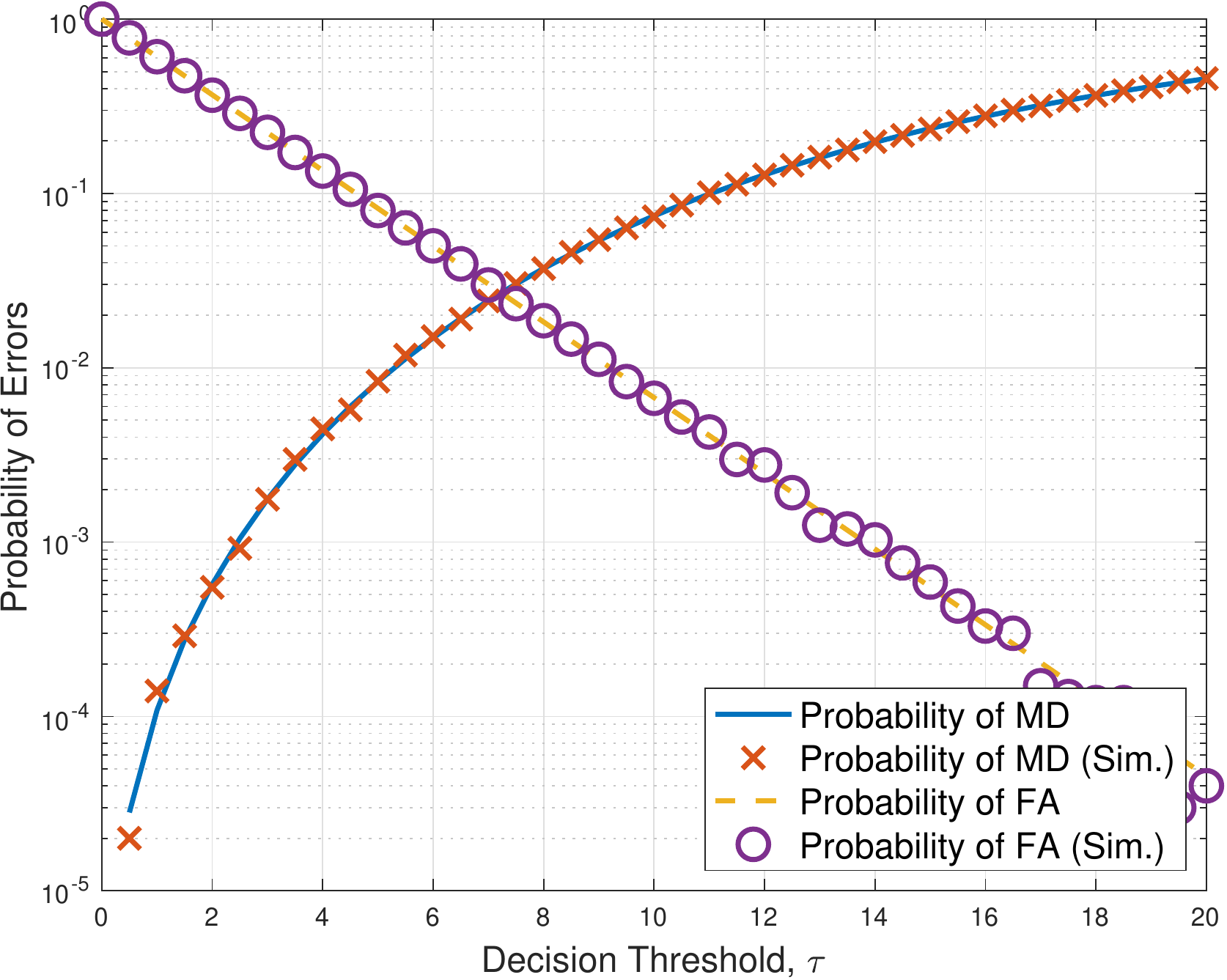}
\end{center}
\caption{The probabilities of MD and FA
as functions of the decision threshold, $\tau$,
when the SNR is 10 dB.}
        \label{Fig:perr_noc}
\end{figure}

For simplicity, we only consider the case that $|\cK_l| = 1$
for $\cH_1$.
Then, under $\cH_1$, $z_l$ is re-written as
\be
z_l = \sqrt{P_{\rm rx}} e^{j \theta_{l(k)}} + n_l.
\ee
For the hypothesis testing, the test statistic becomes
$|z_l|^2$,
which has the following distributions:
\be
\frac{2|z_l|^2}{N_0} \sim
\left\{
\begin{array}{ll}
\chi_2^2, & \mbox{under $\cH_0$} \cr
\chi_2^2 ( \frac{2 P_{\rm rx}}{N_0}), & \mbox{under $\cH_1$,} \cr
\end{array}
\right.
\ee
where $\chi_n^2$ represents
the chi-squared distribution with $n$ degrees 
of freedom and 
$\chi_n^2(\mu)$ represents
the noncentral chi-squared distribution with $n$ degrees 
of freedom and non-centrality parameter $\mu$.
In Fig.~\ref{Fig:perr_noc},
we show the probabilities of MD and FA
as functions of the decision threshold, $\tau$,
when the SNR is 10 dB.
With a proper decision threshold, $\tau$,
we can see that 
both the probabilities of MD and FA can be low.

Consequently, in the rest of the paper,
we assume that the preamble detection can be reliably
carried out, which allows the BS to obtain $\cA$ with negligible
errors.

\section{Multiuser Detection for CDMA Signals}	\label{S:MUD}

In this section, we discuss multiuser detection
for the data transmission phase in the CDMA-based approach.

In the CDMA-based approach,
an active device choosing preamble $l$ transmits a data packet
spread by spreading sequence $l$, $\bc_l$, during the data
transmission phase. Thus,
the received signal during the data transmission
phase at the BS is given by
\begin{align}
\br(t) 
& = 
\sum_{k = 1}^K \bc_{l(k)} h_{k} \sqrt{P_{k}}s_k (t) 
+ \bn_{\rm 2} (t) \in  \uC^{N} \cr
& = 
\sum_{k = 1}^K \bar \bc_{l(k)} \sqrt{P_{\rm rx}} s_k (t) 
+ \bn_{\rm 2} (t),
 \ t = 0, \ldots, D-1,
\end{align}
where $\bar \bc_{l(k)} = \bc_{l(k)} e^{j \theta_k}$
without phase compensation or
$\bar \bc_{l(k)} = \bc_{l(k)}$ with phase compensation,
and $\bn_2 (t) \sim \cC \cN(0, N_0 \bI)$
is the background noise.

Suppose that the ideal outcome of
the preamble detection is obtained. In other words,
the BS knows $\cA$.
From $\cA$, let
\be
\bar \bC = [\bar \bc_{\bar l(1)} \ \cdots \bar \bc_{\bar l(Q)}]
\in \uC^{N \times Q}.
\ee
Then, $\br(t)$ is re-written as
\be
\br (t) = \bar \bC \bx (t)+ \bn_2 (t),
	\label{EQ:rCx}
\ee
where
$\bx(t) = [x_1 (t) \ \ldots \ x_Q (t)]^\rT$
and $x_q (t) = \sum_{k \in \cK_{\bar l(q)}} s_k (t)$.
Using \eqref{EQ:rCx},
the multiuser detection of $Q$ signals can be 
carried out. 
Suppose that $Q \le N$. Then, 
with a sufficiently high SNR,
$\frac{P_{\rm rx}}{N_0}$, $\bx(t)$ can be estimated 
reliably using a multiuser detection approach
including linear detectors 
(such as the minimum mean squared error (MMSE) detector)
\cite{VerduBook} \cite{ChoiJBook2}.
However, among $Q$ signals in $\bx(t)$,
the BS can only detect or decode
the signals transmitted 
from the active devices without preamble collisions. 
Consequently, we expect that the number of successfully detected
or decoded signals is less than or equal to $Q$,
although $Q \le N$.

On the other hand, if $Q > N$, \eqref{EQ:rCx} 
becomes an overloaded system and the 
BS can assume unsuccessful decoding as
detection performance would be poor. 


To see the detection performance when CDMA is used,
we can consider a specific multiuser detector, namely
the MMSE with lattice reduction (MMSE-LR)
detector \cite{ChoiJBook2}.
Suppose that quadrature phase shift keying (QPSK) is employed
for data transmissions during the data transmission phase.
In addition, for spreading sequences in the CDMA-based approach,
we assume Alltop sequences \cite{Alltop80} that
have low cross-correlation.
For a given $N$ (which is a prime greater than or equal to 5),
there are $N^2$ Alltop 
sequences. In each run, 
$L$ out of $N^2$ sequences are randomly selected
for the $L$ spreading sequences, $\{\bc_l\}$.
As mentioned earlier, the MMSE-LR detector is used for
the multiuser detection.

\begin{figure}[thb]
\begin{center}
\includegraphics[width=\figwidth]{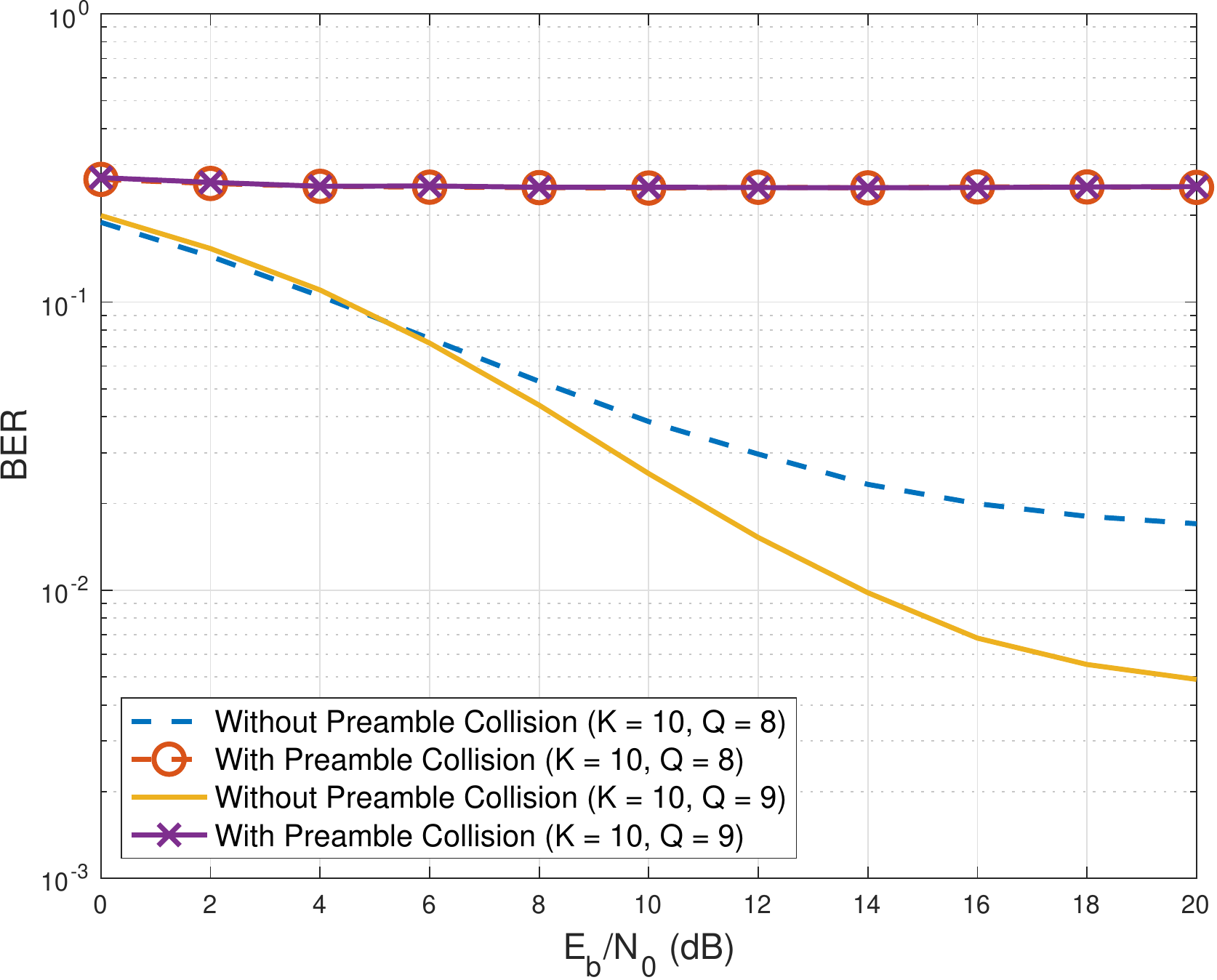}
\end{center}
\caption{BER of MMSE-LR detector as a function of 
$\frac{E_{\rm b}}{N_0}$ 
when $L = 20$, $N = 11$, and $K = 10$.}
        \label{Fig:bplt1}
\end{figure}

Fig.~\ref{Fig:bplt1} shows the bit error
rate (BER) as a function of 
$\frac{E_{\rm b}}{N_0}$,
where $E_{\rm b}$ represents the bit energy,
when $L = 20$, $N = 11$, and $K = 10$.
Since QPSK is used, it is assumed that $E_{\rm b} = P_{\rm rx}/2$. 
With $K = 10$, the following two different cases are considered:
\begin{enumerate}
\item $Q = 8$ and 
4 active devices with preamble collision; 
\item $Q = 9$ and
2 active devices with preamble collision.
\end{enumerate}
From Fig.~\ref{Fig:bplt1},
we can see that a reasonably low BER (less than $10^{-1}$)
is achieved 
for the active devices without preamble collision
when $\frac{E_{\rm b}}{N_0} \ge 6$ dB.
Note that the BER in Fig.~\ref{Fig:bplt1} is uncoded BER. 
Thus, with channel coding, the packet can be reliably decoded.

\begin{figure}[thb]
\begin{center}
\includegraphics[width=\figwidth]{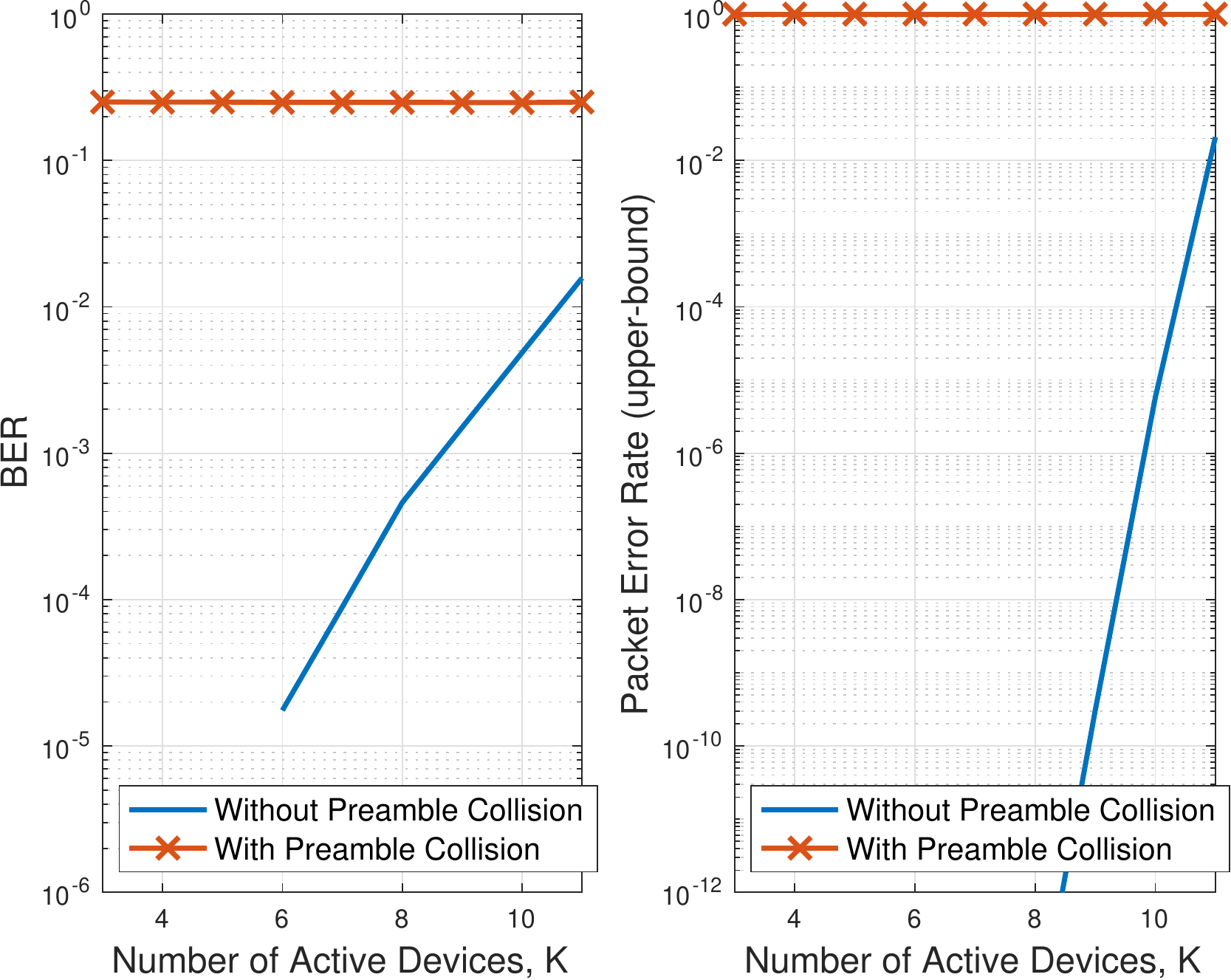} \\
\hskip 0.5cm (a) \hskip 3.5cm (b)
\end{center}
\caption{Performance of MMSE-LR detector as a function of 
the number of active devices, $K$, with one preamble collision 
when $L = 20$, $N = 11$, and $\frac{E_{\rm b}}{N_0} = 20$ dB:
(a) uncoded BER; (b) packet error rate
with BCH code of $(n,k,t) = (255, 191, 8)$.}
        \label{Fig:bplt2}
\end{figure}

Fig.~\ref{Fig:bplt2} shows
the BER and packet error rate as a function
of the  number of active devices, $K$, with one preamble collision 
when $L = 20$, $N = 11$, and $\frac{E_{\rm b}}{N_0} = 20$ dB.
For channel coding, a
Bose-Chaudhuri-Hocquenghem (BCH) code
of $(n,k,t) = (255, 191, 8)$ is considered,
where $n$, $k$, and $t$ represent the code length,
the number of message bits, and the error correction capability,
respectively \cite{WickerBook}.
Since the CDMA-based approach
has multiuser interference, the BER 
(of the active devices without preamble collision) 
increases with $K$.
However, as long as $K$ is smaller than $N$, we see
that both the uncoded BER and packet error rate are low.
Note that in both Figs.~\ref{Fig:bplt1}
and~\ref{Fig:bplt2}, the BER of the active devices
with preamble collision is high as expected,
which means that their data packets cannot be decoded
(as shown in Fig.~\ref{Fig:bplt2} (b)).

Consequently, as long as $K$ is not too larger than $N$,
we can assume that data packets transmitted from the active
devices without preamble collision can be reliably decoded,
while those with preamble collision cannot decoded.
Based on this, the throughput could be 
found as in Section~\ref{S:Thp}.
In the next subsection, we present the simulation
results for the throughput and spectral efficiency.


\section{Throughput and Spectral Efficiency Analysis}	\label{S:Thp}

In this section, we present the main results of performance analysis.
In particular, we derive the throughput, which is defined as
the average number of the active devices
that do not have preamble collisions 
and their packets are successfully decoded if
the condition of $Q \le N$ is met 
for the CDMA-based approach,
and the spectral efficiency.

\subsection{Throughput and
Spectral Efficiency of TDMA}

In this subsection, for comparisons, we consider
the TDMA-based approach, i.e., the case that $N = L$.

For convenience, define
\begin{align}
X_{k,l} =
\left\{
\begin{array}{ll}
1, & \mbox{if active device $k$ chooses preamble $l$} \cr
0, & \mbox{o.w.} \cr
\end{array}
\right.
\end{align}
Let
\be
U_l = \indicator\left( \sum_{k=1}^K X_{k,l} = 1\right)
\in \{0,1\},
\ee
where $\indicator(\cdot)$ represents the indicator function.
Thus, the number of active devices
without preamble collisions becomes $U = \sum_{l=1}^L U_l$.
Clearly, with TDMA for orthogonal channel allocations,
$U$ is also the number of active devices
that can successfully transmit data packets.

If we assume that each active device randomly chooses
one of $L$ preambles,
it can be shown that
\begin{align}
\alpha & = \Pr(U_l = 1) \cr
& = \binom{K}{1} \frac{1}{L}
\left(1 - \frac{1}{L} \right)^{K-1} =
\frac{K}{L} \left(1 - \frac{1}{L} \right)^{K-1}.
\end{align}
Denote by
$\kappa_{\rm td} (K)$
the number of signals that can be decoded with TDMA
for $K$ active devices.
Then, we have
\begin{align}
\kappa_{\rm td} (K) & = \uE[U] = \alpha L 
= K \left(1 - \frac{1}{L} \right)^{K-1}.
\end{align}

To consider the distribution of $K$,
suppose that $M$ devices are uniformly distributed over a unit area.
Furthermore, for a sufficiently large $M$, we assume that
the number of active devices, $K$, follows
a Poisson distribution with mean $\lambda = M p_a$,
where $p_a$ is the access probability.
That is, the average number of active 
devices over the unit area is $\lambda$, which will
be referred to as the activity intensity in the rest of the paper,
and the probability that $K = k$ devices are active\footnote{In fact,
$K$ follows the binomial distribution with parameters
$M$ and $p_a$, i.e.,
$\Pr(K=k) = \binom{M}{k} p_a^k (1-p_a)^{M-k}$. 
However, for a sufficiently large $M$,
\eqref{EQ:Pois} is a good approximation \cite{Mitz05}.}
is
\be
\Pr(K = k) = \frac{\lambda^k}{k!} e^{-\lambda}.
	\label{EQ:Pois}
\ee
Then, the throughput 
becomes
\be
\kappa_{\rm td} = \uE[\kappa_{\rm td} 
(K)] = \lambda e^{-\frac{\lambda}{L}}.
	\label{EQ:k_td}
\ee
In addition, the spectral efficiency,
which is the ratio of the number of 
successfully transmitted data symbols to the length
of slot, is given by
\begin{align}
\eta_{\rm td} 
= \frac{D \kappa_{\rm td}}{T_{\rm slot}}
= \frac{D \lambda e^{-\frac{\lambda}{L}} }{L + L D}.
	\label{EQ:eta_td}
\end{align}
For a fixed $L$, 
as $D \to \infty$, we can see that that
\be
\lim_{D \to \infty}
\eta_{\rm td} = \frac{\lambda}{L} e^{-\frac{\lambda}{L}}
\le e^{-1}.
	\label{EQ:B_tdma}
\ee
In other words, if the overhead of preamble
becomes negligible,
the spectral efficiency is upper-bounded by $e^{-1}$.
However, $D$ may not be large for short message delivery
in MTC. In addition,
the length of slot, $L + LD$, has to be less
than the coherence time for coherent detection at the BS.
Thus, the overhead of preamble has to be taken into account,
which lowers the spectral efficiency.

\subsection{Throughput and
Spectral Efficiency of the Proposed Approach}

In this subsection, we
derive the throughput and spectral efficiency
of the CDMA-based approach.

Prior to finding the throughput, 
we consider the probability of overloaded systems
(i.e., $Q > N$). Let
\begin{align}
W_l = \indicator\left( \sum_{k=1}^K X_{k,l} \ge 2 \right)
\in \{0,1\}.
        \label{EQ:UW_l}
\end{align}
It can be shown that
\begin{align}
\beta & = \Pr(W_l = 1) \cr
& = 1 - \frac{K}{L}
\left(1 - \frac{1}{L} \right)^{K-1} -
\left(1 - \frac{1}{L} \right)^{K}.
	\label{EQ:PWl}
\end{align}
Furthermore, let
$W = \sum_{l=1}^L W_l$.
Recalling that $Q$ 
is the number of unique preambles transmitted,
we have
\be
Q =  U + W.
\ee
Thus, we need to find the distribution of $U+W$,
which is the distribution of the number of bins
with at least 1 ball when 
we throw $K$ balls into $L$ bins uniformly at random \cite{Mitz05}.
Since an exact distribution is not available,
we may resort to an approximation.
To this end, we need the following bound.

\begin{mylemma}
\be
\Pr (U+W \le N) \ge \Pr(W \ge K-N).
	\label{EQ:L1}
\ee
\end{mylemma}
\begin{IEEEproof}
If $U_l = 1$, it implies that
there is only one active device that transmits preamble $l$. In addition,
if $W_l = 1$, the number active devices that transmit preamble $l$
is greater than or equal to 2.
Thus, since there are $K$ active devices,
$\sum_l U_l + 2 \sum_l W_l$ has to be less than or equal to $K$,
which results in 
\be
U + 2 W \le K.
	\label{EQ:U2WK}
\ee
Then, by applying \eqref{EQ:U2WK} to 
$\Pr(U+ W \le N)$, the bound in \eqref{EQ:L1} can be obtained,
which completes the proof.
\end{IEEEproof}

If the $W_l$'s are assumed to be independent (which is not
true, but leads to a good approximation),
we have
\be
\Pr(W \ge K-N\,|\,K > N) 
\approx \sum_{l = K-N}^L \binom{L}{l} \beta^l (1-\beta)^{L-l}.
\ee
Thus, the probability of overloaded systems can be approximately
given by
\be
\Pr(Q > N \,|\, K > N) \approx 
\sum_{l=0}^{K-N-1} \binom{L}{l} \beta^l (1-\beta)^{L-l}.
	\label{EQ:bx}
\ee
Note that if $K \le N$, 
$Q$ cannot be greater than $N$, i.e.,
$\Pr(Q > N \,|\, K \le N) = 0$.
In addition, if $L$ is sufficiently large,
the binomial distribution in 
\eqref{EQ:bx} can be approximated by the Poisson distribution,
which results in the following approximation of
$\Pr(Q  > N\,|\, K > N)$:
\begin{align}
\Pr(Q > N \,|\, K > N) & \approx 
\sum_{l=0}^{K-N-1}  \frac{e^{-\nu} \nu^l}{l!} \cr
& = \frac{\Gamma (K-N, \nu)}{(K-N-1)!}, 
	\label{EQ:aP}
\end{align}
where 
$\nu = L \beta$ and $\Gamma(s,x) = 
\int_x^\infty t^{s-1} e^{-t} dt$ is the upper incomplete Gamma function.

For convenience, let $\kappa_{\rm cd} (K)$
denote the number of signals that can be decoded
in the CDMA-based approach for a given $K$.
If $K \le N$, we have
\begin{align}
\kappa_{\rm cd} (K) & =  \uE[U \,|\, K] =
K \left(1 - \frac{1}{L} \right)^{K-1}.
	\label{EQ:t1}
\end{align}
On the other hand, if $K > N$,
the throughput can be expressed as
\be
\kappa_{\rm cd} (K) = \uE[U \indicator(Q \le N)\,|\, K],
\ee
because multiuser detection fails if $Q > N$.
For a sufficiently large $L$,
the inequality in \eqref{EQ:U2WK}
can be the following
approximation\footnote{If $W_l = 1$,
for a large $L$, it is likely that there are two active
devices that choose preamble $l$. Thus,
the inequality can be replaced with the equality
with a high probability.}:
$U+ 2W  \approx K$.
Thus, $Q \le N$
becomes $W \ge K-N$, which leads to the following approximate 
conditional throughput:
\begin{align}
\kappa_{\rm cd} (K) & \approx \uE[U \indicator(W \ge K-N)\,|\,K] \cr
& \approx \uE[U|K] \Pr(W \ge K-N\,|\,K), \ K > N,
	\label{EQ:t2}
\end{align}
where the second approximation is valid if 
$U$ and $W$ are independent (which is not true).

Note that we have $\Pr(W \ge K-N\,|\,K) = 1$ if $K \le N$
(since $W$ is a non-negative random variable).
Thus, for a given $K$, from \eqref{EQ:t1} and \eqref{EQ:t2}, 
the conditional throughput for any $N \in \{1,\ldots\}$ can be given by
\be
\kappa_{\rm cd} (K) \approx 
\alpha L \Pr(W \ge K-N\,|\,K).
\ee

Then, the average of the approximate throughput 
with \eqref{EQ:t1} can be shown as follows:
\begin{align}
\kappa_{\rm cd} & = \uE[\kappa_{\rm cd} (K)] \cr
& = \sum_{k=0}^N 
\omega_L (k) \Pr(K = k)  \cr
& \ \ + 
\sum_{k=N+1}^\infty  \omega_L (k)
\Pr(Q \le N \,|\, K > N) \Pr(K = k) \cr
& = \sum_{k=0}^\infty \omega_L (k) \Pr(K = k)  \cr
& \ \ -
\sum_{k=N+1}^\infty \omega_L (k)
\Pr(Q > N\,|\, K > N) \Pr(K = k), \ \
	\label{EQ:eta0}
\end{align}
where $\omega_L (k) = k \left(1 - \frac{1}{L} \right)^{k-1}$.
The approximation in \eqref{EQ:aP}
can be used to find a closed-form expression for
the 2nd term on the right-hand side (RHS) in \eqref{EQ:eta0},
which is
\begin{align}
\mbox{2nd term} 
& \approx \sum_{k=N+1}^\infty f(k) \sum_{l=0}^{k-N-1} \psi(l) \cr
& = \sum_{q=0}^\infty \psi(q) \sum_{k = q}^\infty f(N+1+k),
	\label{EQ:2nd}
\end{align}
where
\begin{align*}
f(k) & =  k \left(1 - \frac{1}{L} \right)^{k-1} \Pr(K=k) \cr
\psi(l) & = \frac{e^{-\nu} \nu^l}{l!}.
\end{align*}
In \eqref{EQ:2nd}, since $\sum_q \psi(q) = 1$,
it can be seen that
\begin{align}
\sum_{q=0}^\infty \psi(q) \sum_{k = q}^\infty f(N+1+k)
& = \uE\left[ \sum_{k = Q}^\infty f(N+1+k) \right] \cr
& \approx \sum_{k = \bar Q}^\infty f(N+1+k),
	\label{EQ:2ndx}
\end{align}
where $Q$ is a random variable following the distribution,
$\psi(q)$, and $\bar Q = \uE[Q \,|\, K > N]$.
Substituting  \eqref{EQ:2ndx} into \eqref{EQ:eta0},
we have
\begin{align}
\kappa_{\rm cd} & \approx 
\sum_{k=0}^\infty f(k) -  \sum_{k = \bar Q}^\infty f(N+1+k) 
= \sum_{k=0}^{N+\bar Q} f(k) \cr
& = \sum_{k=0}^{N+\bar Q} \omega_L (k)\Pr(K = k).
\end{align}
Then, for the distribution of $K$ in \eqref{EQ:Pois},
after some manipulations, it can be shown that
\begin{align}
\kappa_{\rm cd} & \approx 
\lambda e^{-\frac{\lambda}{L}}
\underbrace{\frac{\Gamma(N + \bar Q, \bar \lambda)}{(N+ \bar Q - 1)!}}_{=(b)}
\le 
\lambda e^{-\frac{\lambda}{L}} = \kappa_{\rm td},
	\label{EQ:a_eta}
\end{align}
where $\bar \lambda = \lambda \left(1 - \frac{1}{L} \right)$.
The term (b) in \eqref{EQ:a_eta} is the
cumulative distribution function (cdf)
of a Poisson random variable with mean $\bar \lambda$.
Thus, if $N+\bar Q > \bar \lambda$,
we expect that the term (b) can approach 1,
which means that the throughput of
the CDMA-based approach  can be close 
to that of the TDMA-based approach.

Note that since $\nu = L \beta$ and $\beta$ is dependent
on $K$ as shown in \eqref{EQ:PWl},
$\psi (l)$ is also a function of $k$, which is ignored
in \eqref{EQ:2nd}.
That is, the last term in \eqref{EQ:2nd} is another
approximation with a constant $\nu$ or $\beta
= 1 - \frac{K}{L} \left(1 - \frac{1}{L} \right)^{K-1}
- \left(1 - \frac{1}{L} \right)^K$, where $K = N+1$
(which is the smallest $K$ for $\nu$ in \eqref{EQ:aP}).
Thus, in \eqref{EQ:2nd}, $\nu$ is replaced with
\be
\bar \nu = 
L - (N+1) \left(1 - \frac{1}{L} \right)^{N}
- L \left(1 - \frac{1}{L} \right)^{N+1}.
\ee
Thus, we have $\bar Q = \uE[Q \,|\, K > N] = \bar \nu$ in \eqref{EQ:a_eta}.

From \eqref{EQ:a_eta},
the spectral efficiency of the CDMA-based approach is given by
\begin{align}
\eta_{\rm cd} = \frac{\bar D \kappa_{\rm cd}}{T_{\rm slot}}
= \frac{\bar D \lambda e^{-\frac{\lambda}{L}}}{L+N \bar D}
\frac{\Gamma(N + \bar Q, \bar \lambda)}{(N+ \bar Q - 1)!}.
	\label{EQ:eta_cd}
\end{align}
For fixed $L$ and $N$, as $\bar D \to \infty$,
we can see that
\begin{align}
\lim_{\bar D \to \infty}
\eta_{\rm cd} 
= \frac{\lambda e^{-\frac{\lambda}{L}}}{N}
\frac{\Gamma(N + \bar Q, \bar \lambda)}{(N+ \bar Q - 1)!}
\le \frac{L}{N} e^{-1}.
	\label{EQ:B_cdma}
\end{align}
Compared with \eqref{EQ:B_tdma},
we can see that the upper-bound on
the spectral efficiency of the CDMA-based
approach in \eqref{EQ:B_cdma} can be higher than
that of the TDMA-based approach by
a factor of $\frac{L}{N}$.
That is, in terms of the spectral efficiency,
the performance
of the CDMA-based approach can be $\frac{L}{N}$ times better 
than that of the TDMA-based approach.

Without using the upper-bounds
for comparisons between the CDMA-based and TDMA-based approaches,
the ratio of the spectral efficiencies can be directly considered.
From \eqref{EQ:eta_td} and \eqref{EQ:eta_cd},
the performance gain
over the TDMA-based approach with $N = L$
in terms of spectral efficiency can be obtained
as
\begin{align}
\varphi 
& =  \frac{\eta_{\rm cd}}{\eta_{\rm td}} \cr
& \approx
\underbrace{\frac{L}{N} }_{= (a)}
\underbrace{\frac{\Gamma(N + \bar Q, \bar \lambda)}{(N+ \bar Q - 1)!}}_{=(b)},
\ N \le L.
	\label{EQ:ratio}
\end{align}
As mentioned earlier the term (b) is close
to 1 if $N + \bar Q > \bar \lambda$. Since
the term (a)  is greater than or equal to 1,
it is expected that
$\varphi \ge 1$, i.e., 
the CDMA-based approach has a higher spectral efficiency than
the TDMA-based approach.

Note that it is not necessary to consider the case that
$N > L$ (note that if $N = L$, as mentioned earlier, the CDMA-based approach
becomes the TDMA-based approach).
Clearly, $N > L$ is not useful as $N = L$ can avoid
interference in packet transmissions from
the active devices without preamble collisions.

\section{Simulation Results}	\label{S:Sim}

In this section, 
a different set of simulation results are shown
to see the throughput and spectral efficiency.
For simulations, the Poisson arrival is assumed for $K$
unless $K$ is given.


In this subsection, we present the simulation results
of the throughput and spectral efficiency 
of the CDMA-based approach, and compare them
with those of the TDMA-based approach.
All the simulation results are compared
with the theoretical results in 
\eqref{EQ:a_eta}
and \eqref{EQ:eta_cd} for the throughput
and spectral efficiency of the CDMA-based approach,
respectively.

\begin{figure}[thb]
\begin{center}
\includegraphics[width=\figwidth]{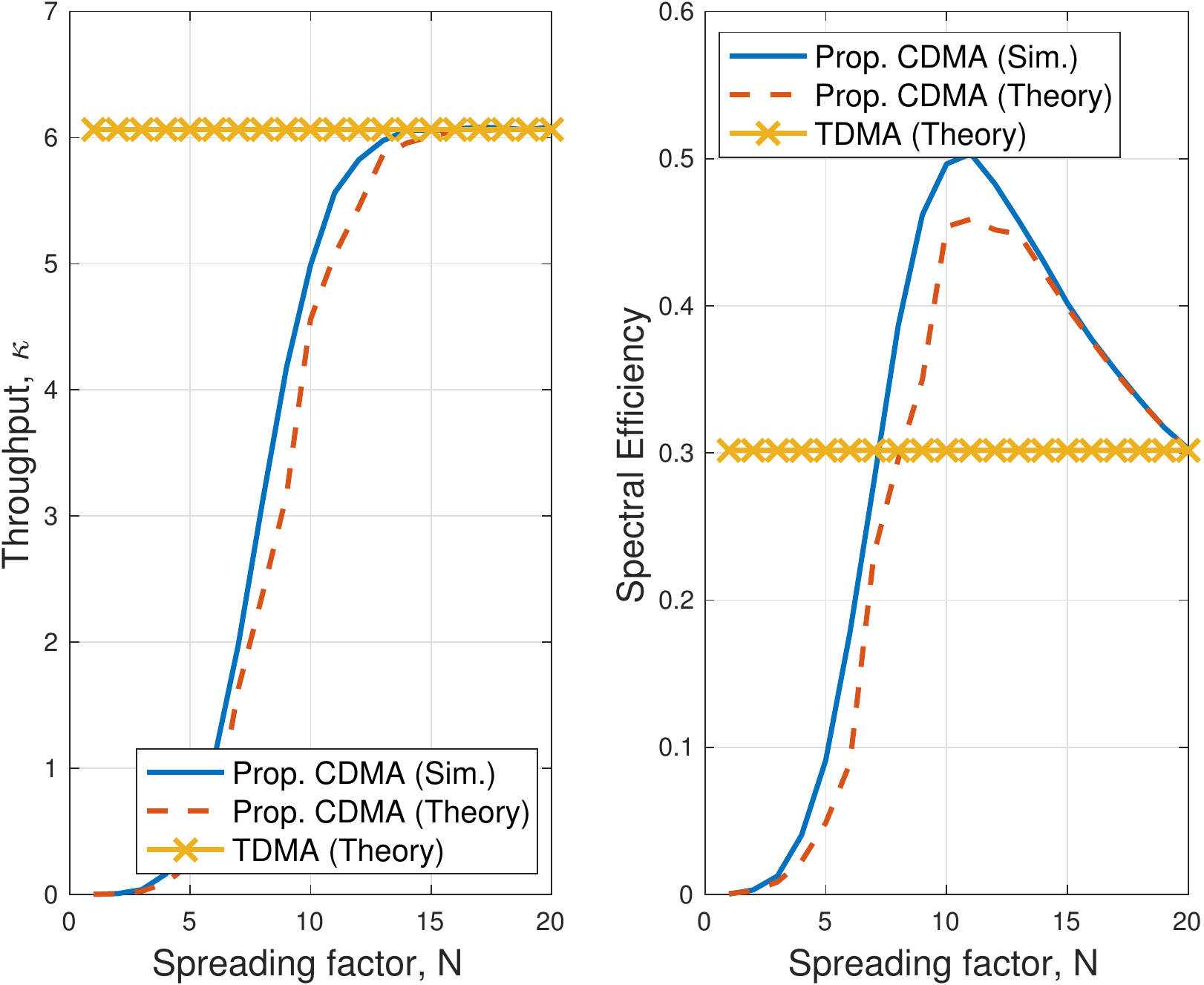} \\
\hskip 0.5cm (a) \hskip 3.5cm (b)
\end{center}
\caption{Performance of the CDMA-based and TDMA-based
approaches in terms of
various values of spreading factor, $N$,
when $L = 20$ and $\lambda = 10$: 
(a) throughput versus $N$; 
(b) spectral efficiency versus $N$ with $D = 200$.} 
        \label{Fig:plt2}
\end{figure}

In Fig.~\ref{Fig:plt2},
we show the throughput
and spectral efficiency of the CDMA-based
approach as functions of spreading factor, $N$,
when $L = 20$ and $\lambda = 10$. 
Note that since the performance
of the TDMA-based approach 
is independent of $N$,
its throughput and spectral efficiency are constants
as shown in Fig.~\ref{Fig:plt2} for comparisons.
In Fig.~\ref{Fig:plt2} (a),
it is shown that the throughput of the CDMA-based
approach can approach that of the TDMA-based approach 
when $N \ge 13$.
Thus, with a shorter length of the data transmission phase
(as $N$ is less than $L$),
it is possible to achieve 
the throughput of the TDMA-based approach,
which results in a higher spectral efficiency.
In Fig.~\ref{Fig:plt2} (b),
it is clearly shown that the spectral efficiency
of the CDMA-based approach can be higher than
that of the TDMA-based approach when $N$ is close to $\lambda$.

\begin{figure}[thb]
\begin{center}
\includegraphics[width=\figwidth]{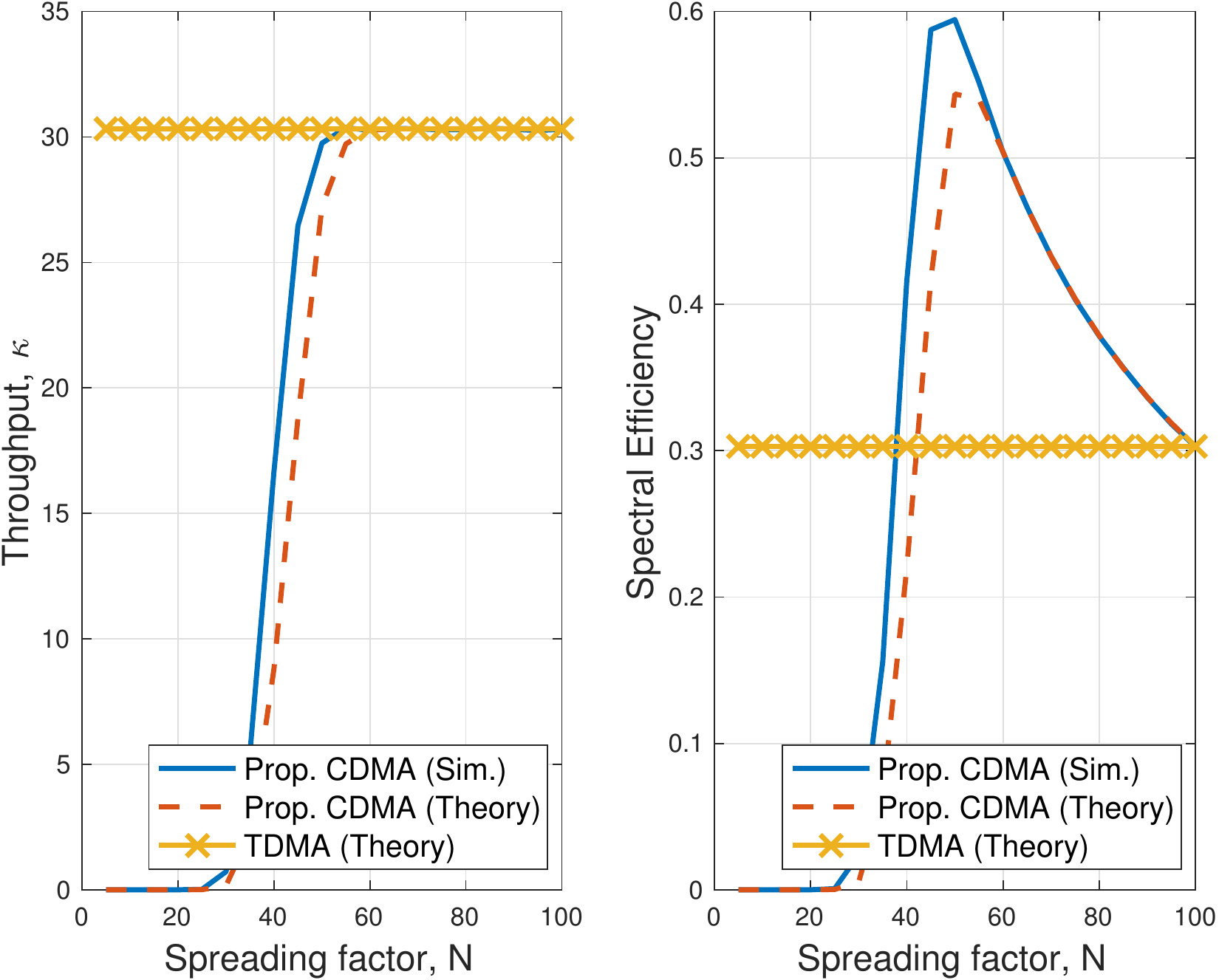} \\
\hskip 0.5cm (a) \hskip 3.5cm (b)
\end{center}
\caption{Performance of the CDMA-based and TDMA-based
approaches in terms of
various values of spreading factor, $N$,
when $L = 100$ and $\lambda = 50$: 
(a) throughput versus $N$;
(b) spectral efficiency versus $N$ with $D = 2000$.}
        \label{Fig:plt2a}
\end{figure}

In Fig.~\ref{Fig:plt2a},
we present simulation results for a larger
system than that used in 
Fig.~\ref{Fig:plt2} by a factor of 5.
That is, we have
$L = 100$ and $\lambda = 50$ in 
Fig.~\ref{Fig:plt2a}.
By comparing Figs.~\ref{Fig:plt2}
and ~\ref{Fig:plt2a},
we can see that the performance gain
of the CDMA-based approach becomes clearer for a larger system.
In addition, in 
Figs.~\ref{Fig:plt2}
and ~\ref{Fig:plt2a},
we can see that the 
theoretical results are reasonably close to simulation results.

\begin{figure}[thb]
\begin{center}
\includegraphics[width=\figwidth]{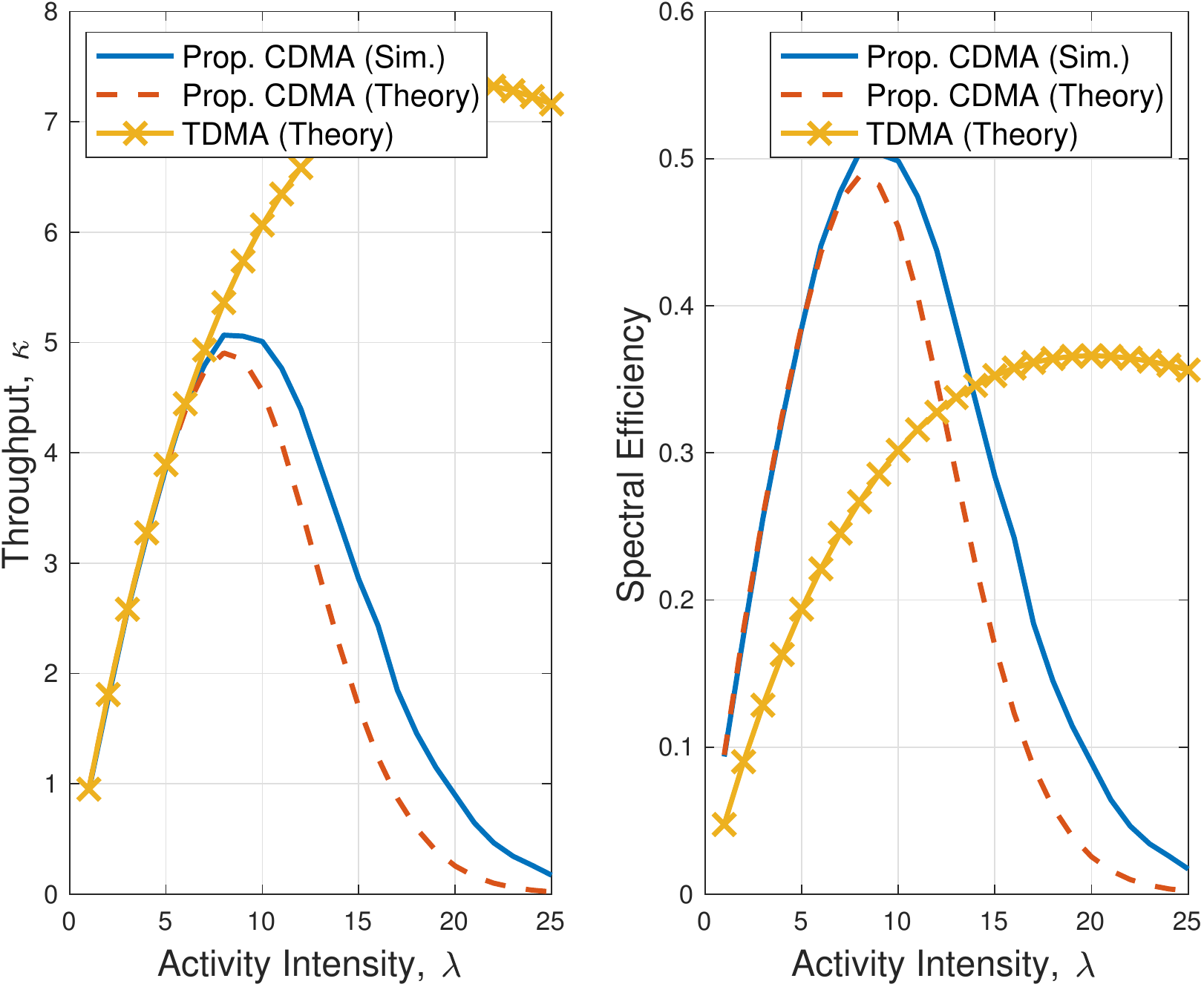} \\
\hskip 0.5cm (a) \hskip 3.5cm (b)
\end{center}
\caption{Performance of the CDMA-based and TDMA-based
approaches in terms of activity intensity, $\lambda$,
when $L = 20$ and $N = 10$: 
(a) throughput versus $\lambda$;
(b) spectral efficiency versus $\lambda$ with $D = 200$.}
        \label{Fig:plt1}
\end{figure}

The throughput and spectral efficiency
are shown as functions of $\lambda$
in Fig.~\ref{Fig:plt1} when $L = 20$ and $N = 10$.
As shown 
in Fig.~\ref{Fig:plt1} (a),
the throughput of the CDMA-based approach
cannot be higher than that of the TDMA-based approach
as $N$ is a half of $L$ for any values of $\lambda$.
However, in Fig.~\ref{Fig:plt1} (b),
it is shown that the spectral efficiency
of the CDMA-based approach can be higher than
the that of the TDMA-based approach.

\begin{figure}[thb]
\begin{center}
\includegraphics[width=\figwidth]{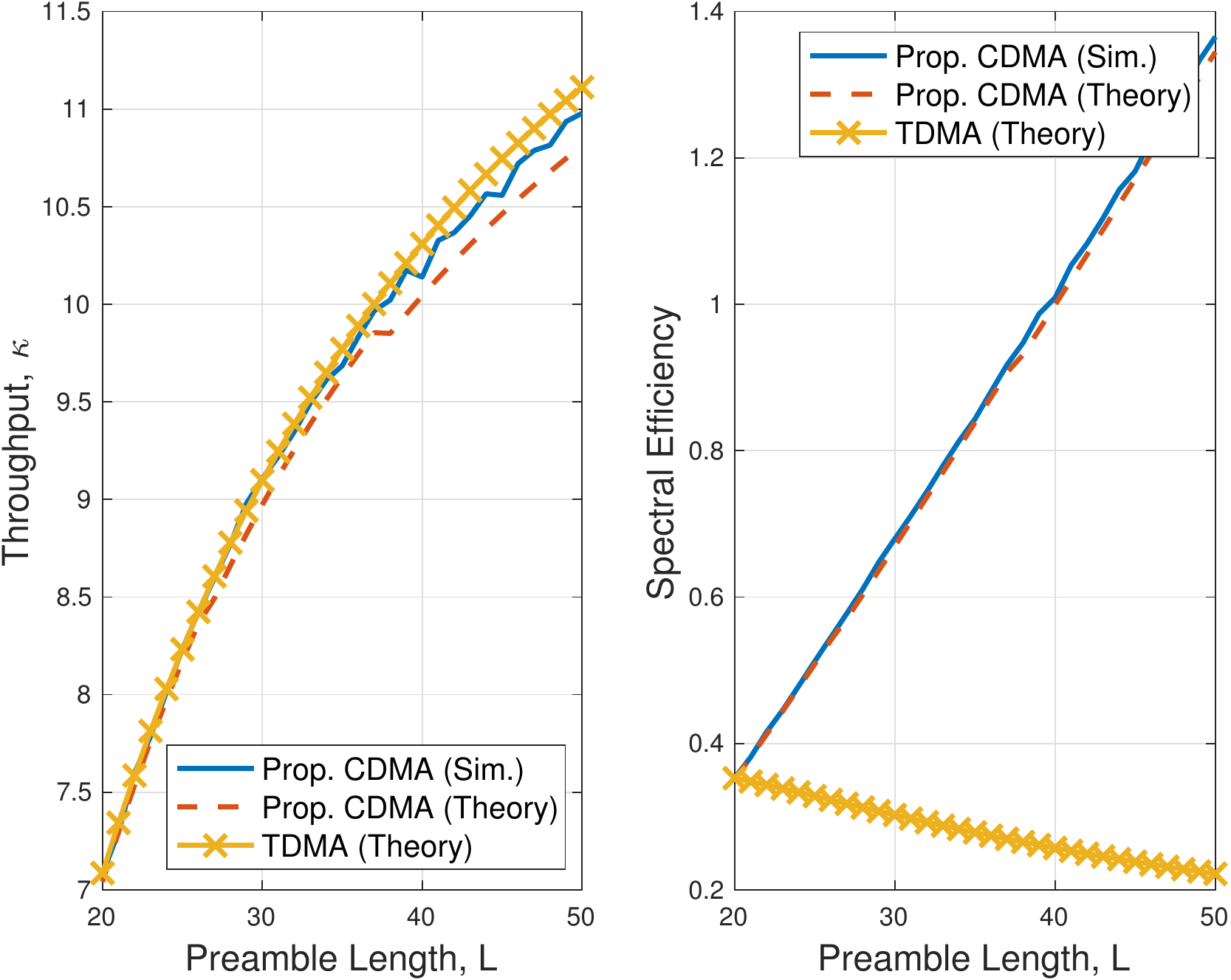} \\
\hskip 0.5cm (a) \hskip 3.5cm (b)
\end{center}
\caption{Performance of the CDMA-based and TDMA-based
approaches in terms of the length of preamble, $L$,
when $N = 20$ and $\lambda = 15$: 
(a) throughput versus $L$;
(b) spectral efficiency versus $L$ with $D = 10 L$.}
        \label{Fig:plt3}
\end{figure}

To see the impact of the length of preamble,
$L$, on the performance in terms of the throughput and
spectral efficiency,
we present results 
when $N = 20$ and $\lambda = 15$
in Fig.~\ref{Fig:plt3}.
As $L$ increases, the probability
of preamble collision decreases, which results
in a higher throughput as 
shown in Fig.~\ref{Fig:plt3} (a).
However, for the TDMA-based approach,
with $D = 10 L$, the length of slot
grows quadratically with $L$.
This leads to the decrease of the spectral efficiency
as shown in Fig.~\ref{Fig:plt3} (b).
On the other hand, 
the spectral efficiency of the CDMA-based approach 
increases with $L$.

\begin{figure}[thb]
\begin{center}
\includegraphics[width=\figwidth]{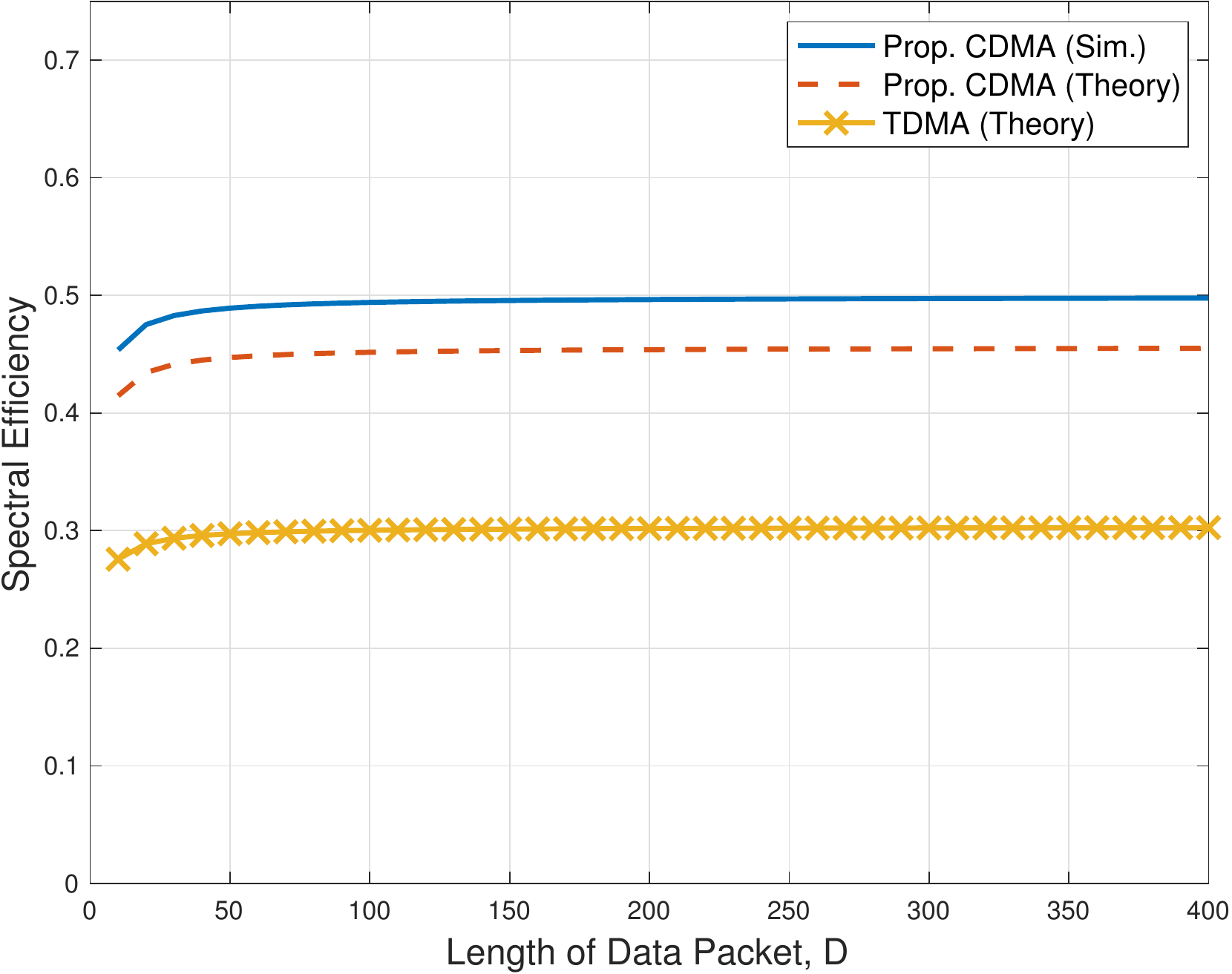}
\end{center}
\caption{Spectral efficiencies of the CDMA-based and TDMA-based
approaches as functions of the length of data packets, $D$,
when $L = 20$, $N = 10$ and $\lambda = 10$.} 
        \label{Fig:plt4}
\end{figure}

Fig.~\ref{Fig:plt4} shows the 
spectral efficiencies of the CDMA-based and TDMA-based
approaches as functions of the length of data packets, $D$,
when $L = 20$, $N = 10$ and $\lambda = 10$.
It is shown that the spectral efficiency increases with $D$, but
gets saturated with a sufficiently large $D$
(e.g., $D \ge 5L$).

\section{Concluding Remarks}	\label{S:Conc}

We studied a 2-step random access
approach that uses CDMA to allow multiple devices
to transmit their data packets simultaneously in this paper. 
Since the spectral efficiency can be improved 
with a spreading factor less than
the number of spreading sequences,
we derived the throughput and then spectral efficiency.
From the analysis, it was shown that
the 2-step random access approach
with CDMA can have a higher spectral efficiency
than the conventional 2-step random access approach
with orthogonal channel allocations
(e.g., TDMA). We also confirmed this with simulation results.

As assumed in the paper, the BS was equipped with a single
antenna. Thus, the space domain was not exploited 
for multiple access. 
Since the space division multiple access (SDMA)
can be combined with CDMA to support more devices,
a generalization to the 
case that the BS is equipped with multiple antenna
would be an interesting further research topic.

\bibliographystyle{ieeetr}
\bibliography{mtc}

\end{document}